\documentclass[pra,aps,amsfonts,amsmath,twocolumn,showpacs,%
  floatfix]{revtex4}
\usepackage{amsfonts}
\usepackage{amsmath}
\usepackage{bm}
\usepackage{epsfig}
\newcommand{\Eq}[1]{Eq.~(\ref{#1})}
\newcommand{\Eqs}[1]{Eqs.~(\ref{#1})}
\newcommand{\Ref}[1]{Ref.~\cite{#1}}
\newcommand{\Refs}[1]{Refs.~\cite{#1}}
\newcommand{\Fig}[1]{Fig.~\ref{#1}}
\newcommand{\Figs}[1]{Figs.~\ref{#1}}
\newcommand{\Sec}[1]{Sec.~\ref{#1}}
\newcommand{\Table}[1]{Table~\ref{#1}}
\renewcommand{\vec}[1]{\bm{#1}}
\newcommand{\beq}{\begin{equation}}
\newcommand{\eeq}{\end{equation}}
\begin{document}
\title{Hartree-Fock-Bogoliubov theory versus local-density
  approximation\\ for superfluid trapped fermionic atoms}
\author{Marcella Grasso}
\author{Michael Urban}
\affiliation{Institut de Physique Nucl\'eaire, F-91406 Orsay Cedex, France}
\begin{abstract}
We investigate a gas of superfluid fermionic atoms trapped in two
hyperfine states by a spherical harmonic potential. We propose a new
regularization method to remove the ultraviolet divergence in the
Hartree-Fock-Bogoliubov equations caused by the use of a zero-range
atom-atom interaction. Compared with a method used in the literature,
our method is simpler and has improved convergence properties. Then we
compare Hartree-Fock-Bogoliubov calculations with the semiclassical
local-density approximation. We observe that for systems containing a
small number of atoms shell effects, which cannot be reproduced by the
semiclassical calculation, are very important. For systems with a
large number of atoms at zero temperature the two calculations are in
quite good agreement, which, however, is deteriorated at non-zero
temperature, especially near the critical temperature. In this case
the different behavior can be explained within the Ginzburg-Landau
theory.
\end{abstract}
\maketitle
\section{Introduction}
In the last few years an increasing interest has been directed towards
ultracold gases of trapped fermionic atoms. Many experimental efforts
are made to develop and improve the techniques for trapping and
cooling fermionic atoms like, for instance, $^{40}$K and $^6$Li. An
interesting aspect of trapped fermionic atoms in comparison with other
Fermi systems is that parameters such as the temperature, the density,
the number of particles, and even the interaction strength are tunable
experimentally. By tuning the magnetic field in the vicinity of a
Feshbach resonance \cite{rob}, the scattering length, which is related
to the interaction strength, can be changed. This offers a wide range
of possibilities to investigate the behaviour of these systems in
different experimental conditions. By using optical or magnetic traps,
temperatures of about $\frac{1}{4} T_F$ have been achieved
\cite{DeMarco,Schreck,Truscott}, where $T_F = \epsilon_F/k_B$ is the
Fermi temperature.

All these efforts are mainly directed to the realization and detection
of a phase transition to the superfluid phase below some critical
temperature $T_C$. In order to have a $s$-wave attractive interaction
among the atoms, which can give rise to $s$-wave pairing correlations
below $T_C$, the atoms have to be trapped and cooled in two different
hyperfine states. This has been achieved in a recent experiment
\cite{OHara}, where also the Feshbach resonance in the $^6$Li
scattering amplitude has been used to enhance the scattering
length. It seems that in the same experiment some signals indicating a
superfluid phase transition have been observed.

From the theoretical point of view many calculations have been
performed in order to predict and study the equilibrium properties of
the trapped system when the phase transition takes place. So far all
these calculations are based on the mean-field approach. In
\Ref{Houbiers} the trapped Fermi gas was treated in local-density
approximation (LDA), where the system is locally treated as infinite
and homogeneous. In \Ref{BaranovPetrov} some corrections to the LDA
for temperatures near $T_C$ were obtained in the framework of the
Ginzburg-Landau (GL) theory. The first approach fully taking into
account the finite system size was introduced in \Ref{brca} and
studied further in \Refs{BruunHeiselberg,BruunPRA66}. It consists in a
Hartree-Fock-Bogoliubov (HFB) calculation, analogous to calculations
done in nuclear physics, where the mean field and the pairing
properties of the system are treated self-consistently. In \Ref{brca}
also a regularization prescription for the pairing field was
developed: Since the densities in the traps are very low, the
atom-atom interaction can be approximated by a zero-range
interaction. However, this leads to an unphysical ultraviolet
divergence of pairing correlations which has to be removed.

In spite of the possibility to perform full HFB calculations, it
should be mentioned that these calculations are numerically very heavy
and therefore limited to moderate numbers of particles. Another
shortcoming of present HFB calculations is that they are restricted to
the case of spherical symmetry, while the traps used in the
experiments are usually strongly deformed. Hence, to describe trapped
systems under realistic conditions, one has to rely on calculations
within the LDA. This is a quite embarrassing situation, since even for
large numbers of particles the results of HFB and LDA calculations
have not always been in good agreement (see results shown in
\Ref{brca}).

In this paper we will present a detailed comparison between HFB and
LDA calculations. In particular, we will show that the disagreement
between HFB and LDA calculations which has been found in \Ref{brca} is
to a certain extent caused by the use of an unsuitable regularization
prescription for the pairing field in the HFB calculations. We will
present a modified regularization prescription which was originally
developed for HFB calculations in nuclear physics \cite{bul} and which
is much easier to implement numerically. (As we learned after sending
the first version of our manuscript, Nygaard et al. used the same
prescription in their calculation of a vortex line in a dilute
superfluid Fermi gas \cite{Nygaard}, without giving a description of
this scheme.) Due to its improved convergence properties, this scheme
leads to more precise results for the pairing field, which in the case
of large numbers of atoms agree rather well with the results of the
LDA at least at zero temperature.  At non-zero temperature, however,
the differences between HFB and LDA results turn out to be important
even for very large numbers of particles. For example, we find that
the critical temperature obtained within the LDA is too high, and that
the pairing field profile near the critical temperature is not well
described by a LDA calculation: we show with the HFB approach that it
actually has a Gaussian shape, as it was predicted in the framework of
the GL theory in \Ref{BaranovPetrov}.

The paper is organized as follows: In \Sec{secformalism} we will
present the adopted formalism with a particular attention on the
description of the regularization techniques. In \Sec{sechfblda} we
will show some comparisons between HFB and LDA calculations and
illustrations of the results obtained with different choices for the
regularization method. We will also discuss results obtained for
non-zero temperatures and verify the quantitative predictions of the
GL theory. Finally, in \Sec{secconclusions} we will draw our
conclusions.

\section{The Formalism}
\label{secformalism}

In this paper we will consider a spherically symmetric harmonic trap
with trapping frequency $\omega$, where $N$ atoms of mass $m$ populate
equally two different spin states $\uparrow$ and $\downarrow$, i.e.,
$N_{\uparrow} = N_{\downarrow}$. As mentioned in the introduction, the
low density of the system allows to introduce a contact interaction
for the atoms, caracterized by the $s$-wave scattering length $a$.
The hamiltonian reads
\begin{equation}
H = T + \sum_{j=1}^N \frac{1}{2} m \omega^2 \vec{r}_j^2
+\frac{4\pi\hbar^2 a}{m} \sum_{i<j} \delta^3
(\vec{r}_i-\vec{r}_j)\,,
\label{1}
\end{equation}
where $T$ is the kinetic term. For convenience let us introduce a
coupling constant $g$ defined as:
\begin{equation}
g = \frac{4 \pi \hbar^2 a}{m}\,.
\label{2}
\end{equation}
Since we are considering attractive interactions, we have $a<0$ and,
consequently, $g<0$. To simplify the notation, we will use in what
follows the ``trap units'', i.e.
\begin{equation}
m = \omega = \hbar = k_B = 1\,.
\label{3}
\end{equation}
Thus, energies will be measured in units of $\hbar \omega$, lengths in
units of the oscillator length $l_{ho} = \sqrt{\hbar/(m\omega)}$, and
temperatures in units of $\hbar\omega/k_B$.

Before describing the HFB approach, let us add some comments on the
validity of the hamiltonian (\ref{1}). The parametrization of the
interaction in terms of the free-space $s$-wave scattering length $a$
is valid at very low densities, where the distance between particles
is much larger than $|a|$. However, if the distance between particles
becomes comparable with $|a|$, the bare interaction has to be replaced
by a density-dependent effective interaction, as it is done in nuclear
physics (see also \cite{Heiselberg}). This is particularly important
in the vicinity of a Feshbach resonance, where $|a|$ becomes very
large. In this case it might be necessary to include the Feshbach
resonance as a new degree of freedom into the Hamiltonian
\cite{Timmermans}.

\subsection{HFB approach and regularization procedure}

The hamiltonian (\ref{1}) will be treated within the mean-field
approximation. We will not go into details here as the formalism has
been introduced and extensively illustrated in \Ref{brca}. The
Hartree-Fock-Bogoliubov (HFB) or Bogoliubov-de Gennes \cite{rs,gen}
equations read:
\begin{equation}
\begin{split}
[ H_0 + W({\vec{R}})] u_{\alpha}({\vec{R}}) +
 \Delta( {\vec{R}} ) v_{\alpha} ({\vec{R}}) 
 &= E_{\alpha} u_{\alpha}({\vec{R}})\,, \\
\Delta({\vec{R}}) u_{\alpha} ({\vec{R}}) - 
[ H_0 + W({\vec{R}}) ]  
v_{\alpha}({\vec{R}}) 
 &= E_{\alpha} v_{\alpha}({\vec{R}}) \,, 
\end{split}
\label{hfbeq}
\end{equation}
where $\alpha$ collects all quantum numbers except spin ($n,l,m$),
$u_\alpha$ and $v_\alpha$ are the two components of the quasiparticle
wavefunction associated to the energy $E_\alpha$, and $H_0$ is the
following single-particle hamiltonian:
\begin{equation}
H_0 = T + U_0 - \mu \,,
\label{h0}
\end{equation}
where $U_0 = \frac{1}{2}r^2$ is the harmonic trapping potential and $\mu$ the
chemical potential. The Hartree-Fock mean field $W(\vec{R})$ in
\Eq{hfbeq} is expressed by
\begin{align}
W({\vec{R}}) = g \sum_{\alpha} \big\{
  &\,|v_{\alpha}({\vec{R}}) |^2\, [1-f(E_{\alpha})]\nonumber\\
  &+|u_{\alpha}({\vec{R}}) |^2 f(E_{\alpha})\big\}\,,
\label{6}
\end{align}
where $f(E_{\alpha})$ is the Fermi function:
\begin{equation}
f(E_{\alpha}) = \frac{1}{e^{E_\alpha/T} +1}\,.
\label{7}
\end{equation}
With a zero-range interaction the pairing field $\Delta(\vec{R})$
appearing in \Eq{hfbeq} would usually be defined as $\Delta({\vec{R}})
= -g \langle \Psi_{\uparrow} (\vec{R}) \Psi_{\downarrow} (\vec{R})
\rangle$, where $\Psi_{\downarrow \uparrow}$ is the field operator
associated with the spin states $\downarrow$ and $\uparrow$. However,
this expression is divergent and must be regularized. The
regularization prescription proposed in \Ref{brca} consists in using
the pseudopotential prescription \cite{huang}:
\begin{equation}
\Delta(\vec{R}) = -g \lim_{r\to0} \frac{\partial}{\partial r}
 \big[ r\, \langle \Psi_{\uparrow} (\vec{R}+\tfrac{\vec{r}}{2})
\Psi_{\downarrow} (\vec{R}-\tfrac{\vec{r}}{2}) \rangle \big] \,.
\label{pseudo}
\end{equation}

In practice, \Eq{pseudo} is evaluated as follows: It is possible to
show that the expectation value $\langle
\Psi_{\uparrow}(\vec{R}+\vec{r}/2)
\Psi_{\downarrow}(\vec{R}-\vec{r}/2)\rangle$ diverges as $\Delta/(4\pi
r)$ when $r \to 0$ if a zero-range interaction is used. Now one adds
and subtracts from this expectation value the quantity $\frac{1}{2}
\Delta(\vec{R}) G^0_{\mu}(\vec{R},\vec{r})$, where $G^0_{\mu}$ is the
Green's function associated to the single-particle hamiltonian $H_0$,
\Eq{h0}, and calculated for the chemical potential $\mu$:
\begin{equation}
G^0_\mu(\vec{R},\vec{r}) = \sum_\alpha
  \frac{\phi^0_\alpha(\vec{R}+\frac{\vec{r}}{2})
    \phi^{0*}_\alpha(\vec{R}-\frac{\vec{r}}{2})}
    {\epsilon^0_\alpha-\mu}\,,
\end{equation}
where $\phi^0_\alpha$ denotes the eigenfunction of $H_0$ with
eigenvalue $\epsilon^0_\alpha-\mu$. One can demonstrate that this
Green's function diverges as $1/(2\pi r)$ when $r\to 0$. Expressing
$\langle\Psi_{\uparrow} \Psi_{\downarrow} \rangle$ in terms of the
wave functions $u$ and $v$, one can write the pairing field $\Delta$
as
\begin{widetext}
\begin{multline}
\Delta(\vec{R}) = -g \lim_{r\to0} \frac{\partial}{\partial r}
\Big[ r \sum_\alpha \Big(
  u_\alpha (\vec{R} + \tfrac{\vec{r}}{2})
  v^*_\alpha (\vec{R} - \tfrac{\vec{r}}{2})\,[1-f(E_\alpha)]
- v^*_\alpha (\vec{R} + \tfrac{\vec{r}}{2})
  u_\alpha (\vec{R} - \tfrac{\vec{r}}{2}) f(E_\alpha)\\
-\frac{\Delta(\vec{R})}{2}\,\frac{
  \phi^0_\alpha(\vec{R}+\tfrac{\vec{r}}{2})
  \phi^{0*}_\alpha(\vec{R}-\tfrac{\vec{r}}{2})}
  {\epsilon^0_\alpha-\mu}\Big)
+\frac{\Delta(\vec{R})}{2} G^0_{\mu} (\vec{R}, \vec{r})\Big]\,.
\label{9}
\end{multline}
\end{widetext}
The sum over $\alpha$ is no longer divergent for $r\to 0$, since the
divergent part of $-\frac{1}{2} \Delta G^0_\mu$ cancels the divergent
part of $\langle \Psi_{\uparrow} \Psi_{\downarrow} \rangle$. Thus, we
can take the limit $r\to 0$ of this sum. On the other hand, the
divergence of the last term is removed by the pseudopotential
prescription, which selects only the regular part of the Green's
function $G^0_\mu$:
\begin{equation}
\lim_{r\to0} \frac{\partial}{\partial r}
  \big[r\, G^0_{\mu} (\vec{R},\vec{r})\big]
  \equiv G_\mu^{0\,\mathit{reg}}(\vec{R})\,.
\label{10}
\end{equation}
Finally, $\Delta$ can be expressed as follows:
\begin{multline}
\Delta({\vec{R}}) = -g \sum_{\alpha} \Big( u_{\alpha} ({\vec{R}}) 
v^* _{\alpha} ({\vec{R}})\, [1-2f(E_{\alpha})]\\
- \frac{\Delta({\vec{R}})}{2}\,
  \frac{|\phi^0_{\alpha} ({\vec{R}})|^2}{\epsilon^0_{\alpha}-\mu} \Big)
- \frac{g \Delta({\vec{R}})}{2} G^{0\,\mathit{reg}}_{\mu}({\vec{R}})\,.
\label{11}
\end{multline}
Once the regular part of the Green's function is calculated for a
given chemical potential $\mu$ \cite{brca}, the HFB equations are
solved self-consistently.

In practice, it is of course impossible to extend the sum over all
states $\alpha$ and one has to introduce some cutoff. However,
since the sum over $\alpha$ converges, the cutoff should not affect
the results if it is chosen sufficiently high. We will discuss about
the rapidity of convergence of the regularization procedure presented
here with respect to the introduced energy cutoff. We will show that
the convergence is quite slow. Moreover, the calculations can become
heavy when systems with a large number of atoms are treated, as the
function $G^{0\,\mathit{reg}}_{\mu}$ has to be calculated for a large
value of the chemical potential. A way to simplify the regularization
procedure and to avoid to calculate $G^{0\,\mathit{reg}}_{\mu}$ is
proposed in \Ref{bul}, where the procedure of \cite{brca} is extended
to calculations for nuclear systems. We will describe this method in
next subsection.

\subsection{Thomas-Fermi approximation in the regularization procedure}
\label{subsectfa}
In \Ref{bul} a simpler regularization procedure was proposed where the
Thomas-Fermi approximation (TFA) is used to calculate the regular part
of the Green's function. To that end let us write the Green's function
$G^0_{\mu}$ by adopting the TFA for the sum over the states
corresponding to oscillator energies $\epsilon^0_{nl}$ above some
suffiently large value $\epsilon_C = N_C+\frac{3}{2}$:
\begin{align}
G^0_{\mu}(\vec{R},\vec{r}) \approx
&\sum_{\substack{nlm\\\epsilon^0_{nl} \le \epsilon_C}}
\frac{\phi^0_{nlm}(\vec{R}+\frac{\vec{r}}{2})
  \phi^{0*}_{nlm}(\vec{R}-\frac{\vec{r}}{2})}
  {\epsilon^0_{nl}-\mu}\nonumber\\
&+ \int_{k_C (R)}^{+\infty}\frac{d^3k}{(2\pi)^3}\,
  \frac{e^{i\vec{k}\cdot\vec{r}}}{\frac{k^2}{2}+\frac{R^2}{2} -\mu} \,,
\label{12}
\end{align}
where
\begin{equation}
k_C(R) = \sqrt{2\epsilon_C-R^2} =
  \sqrt{2N_C+3-R^2} \,.
\label{12bis}
\end{equation}
Observing that
\begin{equation}
\int_0^{+\infty} \frac{d^3k}{(2\pi)^3}\,
  \frac{e^{i\vec{k}\cdot\vec{r}}}{\frac{k^2}{2}} =
  \frac{1}{2\pi r}
\label{13}
\end{equation}
and using \Eq{12}, we can write the regular part of the Green's
function as follows:
\begin{widetext}
\begin{align}
G_\mu^{0\,\mathit{reg}} (\vec{R})
= &\, \lim_{r\to 0} \Big(G^0_{\mu} ( \vec{R}, \vec{r}) - \frac{1}{2\pi r}\Big)
\nonumber\\
\approx& \sum_{\substack{nlm\\\epsilon^0_{nl} \le \epsilon_C}}
\frac{\phi^0_{nlm}(\vec{R})\phi^{0*}_{nlm}(\vec{R})}
  {\epsilon^0_{nl}-\mu}
+ \int_{k_C (R)}^{+\infty}\frac{d^3k}{(2\pi)^3}\,
  \Big(\frac{1}{\frac{k^2}{2}+\frac{R^2}{2} -\mu}
    -\frac{1}{\frac{k^2}{2}}\Big)
-\int_0^{k_C(R)} \frac{d^3k}{(2\pi)^3}\,
  \frac{1}{\frac{k^2}{2}}\,.
\label{gregtf}
\end{align}
Evaluating the integrals over $\vec{k}$ and summing over the magnetic
quantum number $m$, we obtain
\begin{equation}
G_{\mu}^{0\,\mathit{reg}} (\vec{r}) \approx
\sum_{\substack{nl\\\epsilon^0_{nl} \le \epsilon_C}}
\frac{(2l+1) R_{nl}^2(r)}{4\pi(\epsilon^0_{nl}-\mu)}
+ \frac{k^0_F(r)}{2 \pi^2} \ln \frac{k_C(r)+k^0_F(r)}{k_C(r)-k^0_F(r)}
- \frac{k_C(r)}{\pi^2}\,,
\label{16}
\end{equation}
\end{widetext}
where $R_{nl}$ are the radial parts of the oscillator wave functions and
\begin{equation}
k^0_F(r) = \sqrt{2\mu-r^2}
\label{kf0}
\end{equation}
is the local Fermi momentum. As noted in \Ref{bul}, this method can be
used beyond the classical turning point (characterized by $k^0_F(r) =
0$) by allowing for imaginary values of $k^0_F(r)$. The case that
$k_C(r)$ becomes imaginary will not be considered, because we assume
that $N_C$ is sufficiently large such that the pairing field can be
neglected in the regions where $k_C(r)$ is imaginary. It should also
be pointed out that already for, say, $N_C \ge \mu+10$, \Eq{16} is an
extremely accurate approximation to $G_\mu^{0\,\mathit{reg}}$, and
gives results which are almost undistinguishible from those obtained
by the numerically heavy algorithm proposed in \Ref{brca}.

Now let us substitute \Eq{16} into \Eq{11}. We have to choose a cutoff
for the sum over single-particle states. Instead of choosing a cutoff
for the quasiparticle energies $E_\alpha$, as it is done in
Ref. \cite{bul}, we can likewise restrict the sum in \Eq{11} to the
states corresponding to those appearing in the sum in \Eq{16}. This is
the natural choice if one obtains the wave-functions $u_\alpha$ and
$v_\alpha$ and the quasiparticle energies $E_\alpha$ by solving
\Eq{hfbeq} in a truncated harmonic oscillator basis containing the
states satisfying $\epsilon_{nl}^0 \le \epsilon_C =
N_C+\frac{3}{2}$. In this way we obtain the following simple formula
for the gap:
\begin{widetext}
\begin{equation}
\Delta(r)= -g 
\sum_{\substack{nl\\\epsilon^0_{nl} \le \epsilon_C}}
 \frac{2l+1}{4\pi}u_{nl}(r) v_{nl} (r)\, [1-2f(E_{nl})]
-g \frac{\Delta(r)}{2} \Big(\frac{k^0_F(r)}{2\pi^2} \ln
  \frac{k_C(r) + k^0_F(r)}{ k_C (r) - k^0_F (r)} - \frac{k_C(r)}{\pi^2}
  \Big)\,.
\label{18}
\end{equation}
Finally, this can be rewritten in terms of a position dependent
effective coupling constant:
\begin{equation}
\Delta(r) = -g_{\mathit{eff}} (r) 
\sum_{\substack{nl\\\epsilon^0_{nl} \le \epsilon_C}}
\frac{2l+1}{4\pi}u_{nl}(r) v_{nl} (r)\, [1-2f(E_{nl})]\,,
\label{19}
\end{equation}
where
\begin{equation}
\frac{1}{g_{\mathit{eff}}(r)} = \frac{1}{g} + \frac{1}{2\pi^2} 
\Big(\frac{k^0_F(r)}{2}
  \ln \frac{k_C (r) + k^0_F (r)}{k_C (r) - k^0_F (r)} - k_C(r) \Big)\,.
\label{20}
\end{equation}
\end{widetext}
We stress again that the results obtained with this regularization
prescription, from now on called prescription (a), coincide with the
results obtained with the prescription introduced in \Ref{brca}.

However, it will turn out that it is useful to introduce the following
modification of the method: Let us replace everywhere $k^0_F(r)$ by
the local Fermi momentum taking into account the full potential
(trapping potential $U_0$ plus Hartree-Fock potential $W$):
\begin{equation}
k_F(r) = \sqrt{2\mu-r^2-2W(r)}\,.
\label{kf}
\end{equation}
Formally this replacement does not change anything: Instead of adding
and subtracting the term $\frac{1}{2} \Delta(\vec{R})
G^0_\mu(\vec{R},\vec{r})$ from the divergent expectation value
$\langle \Psi_{\uparrow} (\vec{R} + \vec{r}/2) \Psi_{\downarrow}
(\vec{R} - \vec{r}/2) \rangle$ with $G_\mu^0$ being the Green's
function corresponding to the harmonic oscillator potential $U_0$, we
can also add and subtract a similar term involving the Green's
function $G_\mu$ corresponding to the full potential $U_0+W$. Also
from \Eq{20} it is evident that in the limit $N_C\to\infty$ [i.e.,
$k_C(r)\to\infty$] the results will be independent of the choice of
$k_F$. However, we will see that the convergence of this modified
scheme, from now on referred to as scheme (b), is very much
improved. Thus, it is possible to use a much smaller cutoff $N_C$
without having a strong cutoff dependence of the results.
\subsection{Local-density approximation}
\label{subseclda}
If the number of particles becomes very large, it is natural to assume
that the system can be treated locally as infinite matter with a local
chemical potential given by $\mu - U_0(\vec{r})$. This assumption
leads directly to the local-density approximation (LDA). Formally, the
LDA corresponds to the leading order of the Wigner-Kirkwood $\hbar$
expansion, which is at the same time an expansion in the gradients of
the potential \cite{rs}. Thus it is the generalization of the standard
Thomas-Fermi approximation (TFA), which also corresponds to the
leading order of an $\hbar$ or gradient expansion, to the superfluid
phase. Here we will adopt the name LDA in order to avoid confusion
with the full HFB calculations using the TFA only in the
regularization prescription, as discussed in \Sec{subsectfa}. But in
the literature also the name TFA is adopted.

In the case of a zero-range interaction, the LDA (or TFA) amounts to
solving at each point $\vec{r}$ the following non-linear equations for
the mean field $W(\vec{r})$ and the pairing field $\Delta(\vec{r})$:
\begin{widetext}
\begin{gather}
W(\vec{r}) = \frac{g}{2}\rho(\vec{r}) = g \int
  \frac{d^3k}{(2\pi)^3}\Big(\frac{1}{2}
  -[1-2f(E(\vec{r},\vec{k}))]
    \frac{\epsilon(\vec{r},\vec{k})-\mu}{2E(\vec{r},\vec{k})}\Big)~
\label{lda1},\\
\Delta(\vec{r}) = -g \int \frac{d^3k}{(2\pi)^3}
  \Big([1-2f(E(\vec{r},\vec{k}))]
  \frac{\Delta(\vec{r})}{2E(\vec{r},\vec{k})}
  -\frac{\Delta(\vec{r})}{2(\epsilon(\vec{r},\vec{k})-\mu)}\Big)\,,
\label{lda2}
\end{gather}
\end{widetext}
where
\begin{gather}
\epsilon(\vec{r},\vec{k}) =
  \frac{k^2}{2}+U_0(\vec{r})+W(\vec{r})\,,
\label{lda3a}\\
E(\vec{r},\vec{k}) =
  \sqrt{(\epsilon(\vec{r},\vec{k})-\mu)^2+\Delta^2(\vec{r})}\,.
\label{lda3b}
\end{gather}
The last term in \Eq{lda2} has been introduced in order to regularize
the ultraviolet divergence. In fact, the pseudopotential prescription
used in the previous subsections was originally motivated by the fact
that it reduces to such a term if it is applied to a homogeneous
system \cite{brca,bul}. A more rigorous justification of this term is
that it appears if one renormalizes the scattering amplitude of two
particles in free space \cite{Melo}.

Let us first consider the case of zero temperature, $T = 0$. In this
case, and if the gap $\Delta$ is small compared with the local Fermi
energy $\epsilon_F = \frac{1}{2}k_F^2$, \Eqs{lda1} and (\ref{lda2})
can be solved (almost) analytically. Under these conditions the
density practically coincides with the density obtained for $\Delta =
0$, where \Eqs{lda1}, (\ref{lda3a}), and (\ref{lda3b}) can be
transformed into a cubic equation for the local Fermi momentum:
\begin{equation}
g\frac{k_F^3(\vec{r})}{6\pi^2}
  +\frac{k_F^2(\vec{r})}{2}+U_0(\vec{r})-\mu = 0\,.
\end{equation}
For a given local Fermi momentum and under the assumption that
corrections of higher order in $\Delta/\epsilon_F$ are negligible,
\Eq{lda2} can be solved analytically. The result is the well-known
formula
\begin{equation}
\Delta(\vec{r}) = 8\epsilon_F(\vec{r})
  \exp\Big(-2-\frac{\pi}{2k_F(\vec{r})|a|}\Big)\,.
\label{deltaanalytic}
\end{equation}

Now we turn to the case of non-zero temperature, but we want to
consider only temperatures below the critical temperature, i.e.,
$0<T<T_C$. Therefore, we can neglect the influence of the temperature
on the density and have to consider only the temperature dependence of
$\Delta$. Let us denote the gap at $T = 0$ by $\Delta_0$. Then the gap
at non-zero temperature can be obtained from the approximate relation
\cite{Lifshitz}
\begin{equation}
-\ln\frac{\Delta(\vec{r})}{\Delta_0(\vec{r})}
  = \int\! d\xi\, \frac{f\big(\sqrt{\xi^2+\Delta^2(\vec{r})}\big)}
    {\sqrt{\xi^2+\Delta^2(\vec{r})}}\,.
\end{equation}
The solution of this equation leads to a universal function which
gives the ratio $\Delta/\Delta_0$ as a function of $T/T_C$, with $T_C
\approx 0.57 \Delta_0$. Note that, within the LDA, the critical
temperature is a local quantity, $T_C = T_C(\vec{r})$.

In order to compare the LDA with the HFB theory, with special emphasis
on the regularization prescription, we will now introduce a
regularization scheme for the gap equation within LDA which is
slightly different from \Eq{lda2}. First of all, if we want to
investigate the cutoff dependence, we have to introduce a cutoff in
\Eq{lda2}. Secondly, the regularization term introduced in \Eq{lda2}
corresponds to the regularization prescription (b) described at the
end of the previous subsection, which is different from that
introduced in \Ref{brca} and from the regularization scheme (a). If we
want to compare the LDA results with HFB results obtained with the
original prescription or with the prescription (a), which involves the
Green's function $G^0_\mu$ of the potential $U_0$ and not the Green's
function $G_\mu$ of the full potential $U_0+W$, we have to replace the
energy $\epsilon(\vec{r},\vec{k})$ appearing in the regularization
term by
\begin{equation}
\epsilon^0(\vec{r},\vec{k}) = \frac{k^2}{2}+U_0(\vec{r})\,.
\end{equation}
Thus, the gap equation within LDA suitable for comparison with the
regularization scheme (a) reads
\begin{widetext}
\begin{equation}
\Delta(\vec{r}) = -g \int_0^{k_C(\vec{r})} \frac{d^3k}{(2\pi)^3}
  \Big([1-2f(E(\vec{r},\vec{k}))]
  \frac{\Delta(\vec{r})}{2E(\vec{r},\vec{k})}
  -\frac{\Delta(\vec{r})}{2(\epsilon^0(\vec{r},\vec{k})-\mu)}\Big)\,.
\end{equation}
At zero temperature, $T = 0$, it is again possible to solve this
equation analytically, with the result
\begin{equation}
\Delta(\vec{r}) = 8\epsilon_F(\vec{r})
  \sqrt{\frac{k_C(\vec{r})-k_F(\vec{r})}{k_C(\vec{r})+k_F(\vec{r})}}
\exp\Big(-2-\frac{\pi}{2k_F(\vec{r})|a|}-
  \frac{k^0_F(\vec{r})}{2k_F(\vec{r})}
  \ln\frac{k_C(\vec{r})-k^0_F(\vec{r})}{k_C(\vec{r})+k^0_F(\vec{r})}\Big)\,.
\label{ldacut}
\end{equation}
\end{widetext}
The result corresponding to the regularization scheme (b),
\Eq{deltaanalytic}, is recovered from this result by replacing $k^0_F$
by $k_F$. In this case there is no cutoff dependence at all, but one
should remember that in deriving \Eq{ldacut} we have implicitly
assumed that the cutoff lies above the Fermi surface. A weak cutoff
dependence would appear only if corrections to \Eq{ldacut} of higher
order in $\Delta/\epsilon_F$ were included.
\section{Numerical results}
\label{sechfblda}
In this section we will present some numerical results. In particular, 
we will investigate the convergence properties of the different
renormalization methods. Then, we will discuss the validity of the LDA
at zero temperature. Finally, we will compare HFB and LDA calculations
at non-zero temperature.

In our numerical calculations we will use for the coupling constant
the value $g = -1$ (in units of $\hbar^2 l_{ho}/m$). If we consider
$^6$Li atoms with scattering length $a = -2160 a_0$ \cite{Abraham},
where $a_0 = 0.53\, \mbox{\AA}$ is the Bohr radius, this value of $g$
corresponds to a trap with $\omega = 2\pi\times
817\,\mbox{Hz}$. (Before relating this to real experimental
conditions, one should however remember that in the experiments the
trap is usually axially deformed, with a low longitudinal trapping
frequency $\omega_z$ and a high transverse trapping frequency
$\omega_\perp$. For example, in the experiment described in
\Ref{OHara}, the trapping frequencies were given by $\omega_z =
2\pi\times 230\,\mbox{Hz}$ and $\omega_\perp = 2\pi\times
6625\,\mbox{Hz}$.) The choice $g = -1$ also facilitates the comparison
of our results with those from \Ref{brca}, where the same value for
$g$ was used.

\subsection{Convergence of the regularization methods}
In this section we will discuss the convergence rates with respect to
the cutoff used in the numerical calculations for different choices
for the regularization procedure. As in \Sec{secformalism} we denote
by (a) the HFB calculations made with the choice of $k^0_F$ given by
\Eq{kf0}, and and by (b) the calculations made with the choice where
$k^0_F$ is replaced by $k_F$ as given by \Eq{kf}. For our comparison
we use a chemical potential $\mu = 32\hbar \omega$, the corresponding
number of atoms in the trap is $N\approx 1.7\times 10^4$.

\begin{figure}
\begin{center}
\epsfig{file=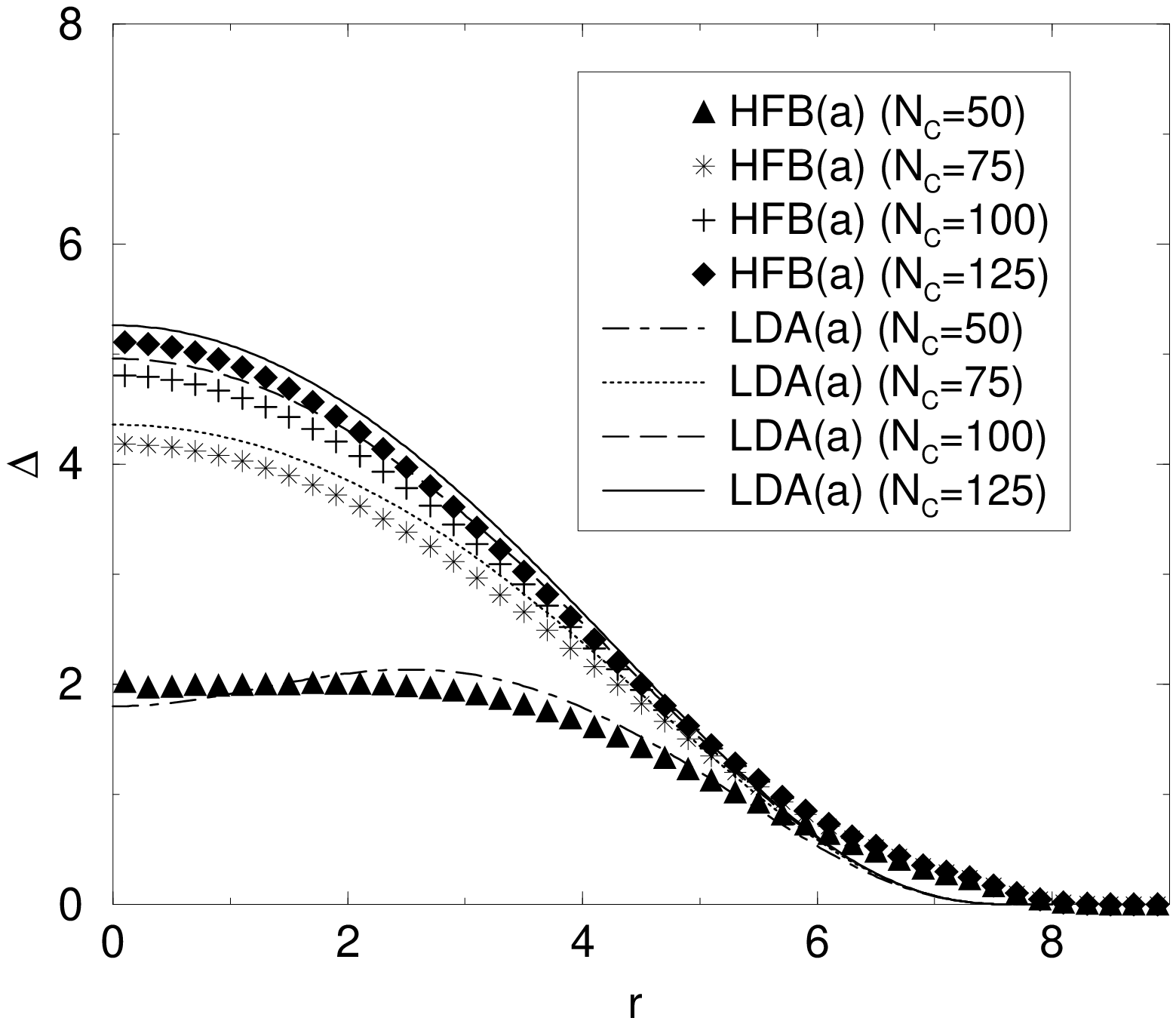,height=6.2cm}
\end{center}
\caption{Pairing field $\Delta$ (in units of $\hbar \omega$) 
as a function of the distance $r$ (in units of $l_{ho}$) from
the center of the trap, calculated for the parameters
$\mu = 32\hbar\omega$ and $g = -1 \hbar^2 l_{ho}/m$, corresponding to
$N\approx 1.7\times 10^4$ particles in the trap. The different curves
have been obtained within the HFB and LDA formalisms using the
regularization prescription (a) for different values of the cutoff
$N_C$.\label{fig1}}
\begin{center}
\epsfig{file=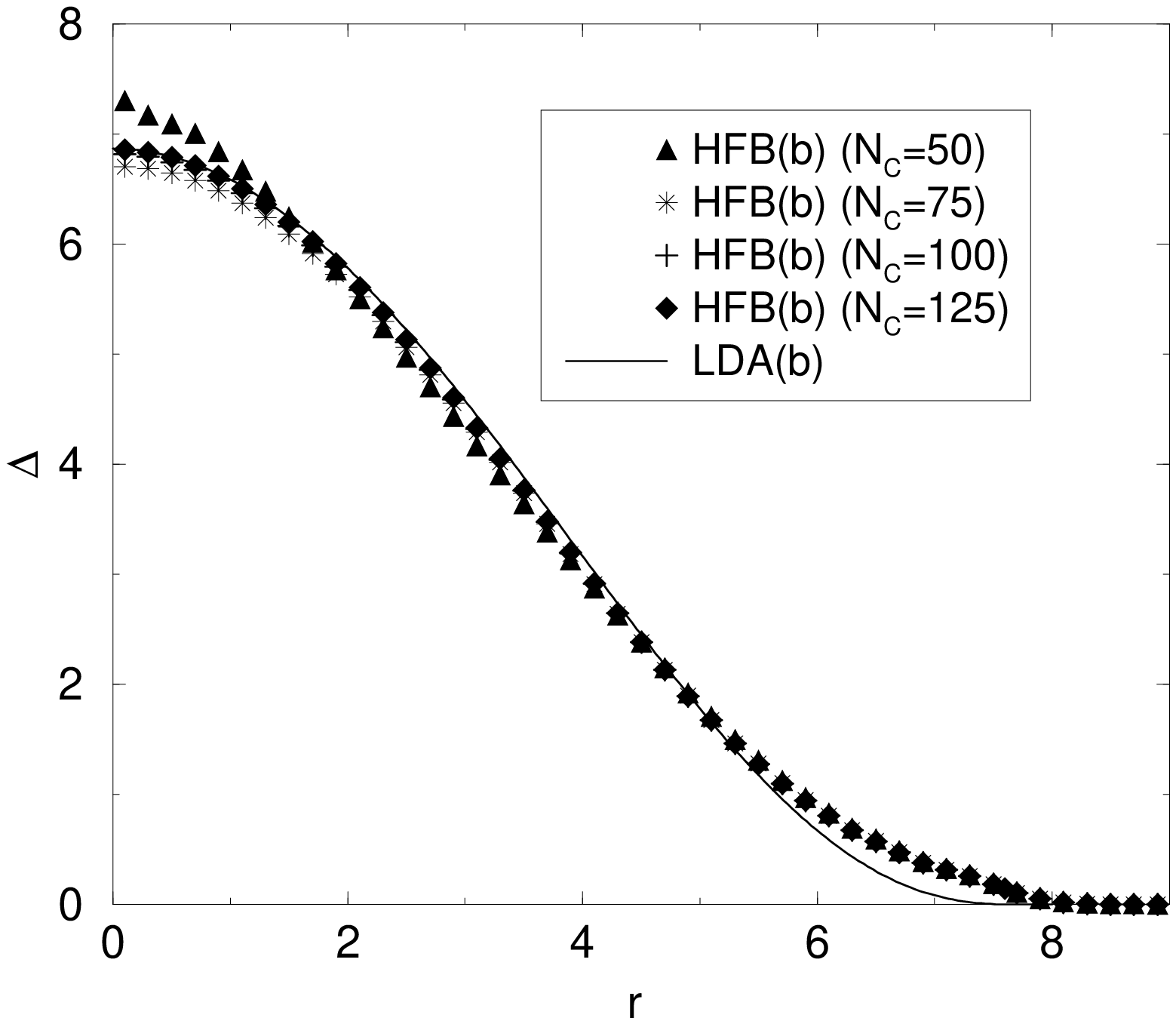,height=6.2cm}
\end{center}
\caption{Same as \Fig{fig1}, but with regularization prescription
(b). Remember that with this prescription the LDA result [\Eq{ldacut}]
is independent of the cutoff $N_C$. \label{fig2}}
\end{figure}

In \Figs{fig1} and \ref{fig2} we present the pairing field $\Delta$
calculated at zero temperature within the HFB and LDA formalisms for
different values of the cutoff $N_C$ from $50$ up to $125$. The
results shown in \Fig{fig1} have been obtained with the choice (a) for
the regularization for both the HFB and LDA calculations. We verified
that the HFB calculations with the exact Green's function
$G^{0\,\mathit{reg}}_\mu$ (without TFA) give practically the same
results as the method HFB(a) for all the values of the cutoff. This
means that the TFA in the regularization procedure is very satisfying
and reproduces well the regular part of the oscillator Green's
function.

We observe in \Fig{fig1} that the agreement between LDA and HFB is
reasonable for all values of the cutoff $N_C$. We also notice that for
$N_C=125$, which is the maximum value that we considered, the
convergence has not yet been reached and therefore the pairing field
would grow further if we could increase the cutoff above 125.
In \Fig{fig2} we present the same calculations made with the choice
(b) for the regularization. Remember that with this choice, the
pairing field within LDA is independent of $N_C$ once $N_C$ lies above
the Fermi surface. On the other hand, the HFB results saturate quite
fast and are already very close to convergence for $N_C=75$. Again,
the LDA and HFB results are in reasonable agreement.

\begin{figure}
\begin{center}
\epsfig{file=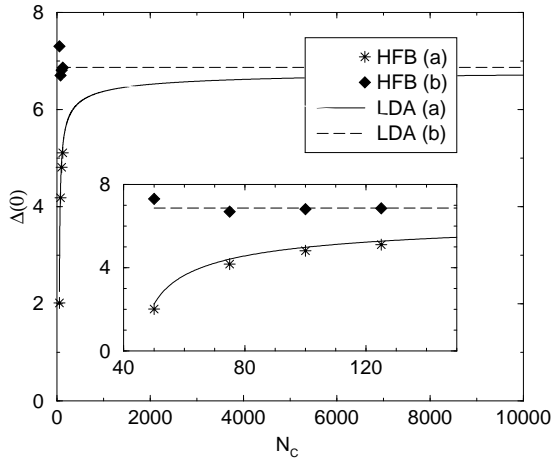,height=6.2cm}
\end{center}
\caption{Value of the pairing field in the center of the trap,
$\Delta(0)$ (in units of $\hbar \omega$), 
as a function of the cutoff $N_C$, obtained from
HFB calculations with the regularization methods (a) (stars)
and (b) (diamonds), and from the LDA, method (a) (solid line) and
method (b) (dashed line). The parameters $\mu$ and $g$ are the same as
in \Fig{fig1}.\label{fig3}}
\end{figure}

By comparing \Figs{fig1} and \ref{fig2} one observes clearly that the
calculations (a), \Fig{fig1}, are still quite far from convergence
even for the highest considered cutoff. We argue that the convergence
rate of method (a), which is the same convergence rate as that of HFB
without TFA in the regularization prescription \cite{brca}, is much
slower than that of method (b). This is more evident in \Fig{fig3}
where we plot the HFB values of the pairing field in the center of the
trap, $\Delta(0)$, for the two regularization prescriptions (a)
(stars) and (b) (diamonds) as a function of the cutoff $N_C$. We also
plot the results obtained within the LDA(a) (full line) and LDA(b)
(dashed line) up to a cutoff of $N_C=10^4$. In the inset of the figure
we magnify the region of cutoff values between 50 and 150. We can
observe in the inset that the LDA(a) curve fits well the calculated
points for HFB(a). We noticed that the LDA(a) results converge slowly
towards a pairing field of about $6.86 \hbar \omega$, at a very high
cutoff, $N_C = 10^6$. For $N_C = 10^3$ the pairing field in LDA(a) is
still only $6.37 \hbar \omega$. This very slow convergence rate can be
understood within the LDA by taking the ratio of the pairing fields
corresponding to the methods (a) and (b). Using \Eq{ldacut} in the
limit of very large $k_C$, one can derive the relation
\begin{equation}
\frac{\Delta_{\mathit{LDA(a)}}(\vec{r})}{\Delta_{\mathit{LDA(b)}}(\vec{r})}
  \approx 1 - \frac{|g| \sqrt{2}\,[\mu-W(\vec{r})]}
    {3 \pi^2 \sqrt{N_C}} + \cdots\,,
\label{22}
\end{equation}
where $W(\vec{r})$ represents the Hartree field (in the present case,
$W(0)\approx -16\hbar\omega$).

As the agreement between LDA(a) and HFB(a) is good in the region up to
$N_C=125$, we suppose that the convergence rate for HFB(a) is the same
as for LDA(a). On the contrary, within HFB(b) the values of the
pairing field in the center of the trap are $6.81 \hbar \omega$ for
$N_C = 100$ and $6.86 \hbar \omega$ for $N_C = 125$: we conclude that
the convergence in this case is much faster. In what follows we will
always use the method (b) for the regularization procedure.

\subsection{Validity of the LDA at zero temperature}
As mentioned before, the parameters used for the calculations shown in
\Figs{fig1}, \ref{fig2}, and \ref{fig3} correspond to a trap with
about $1.7 \times 10^4$ atoms. In this case we found a good agreement
between the numerical HFB results and the results obtained from the
LDA. However, one might wonder under which conditions the LDA is
valid. To study this question, one has to look at systems containing
smaller numbers of particles, since in smaller systems the quantum
effects (in particular shell effects) which are neglected in the LDA,
are supposed to be more important.

\begin{figure}
\begin{center}
\epsfig{file=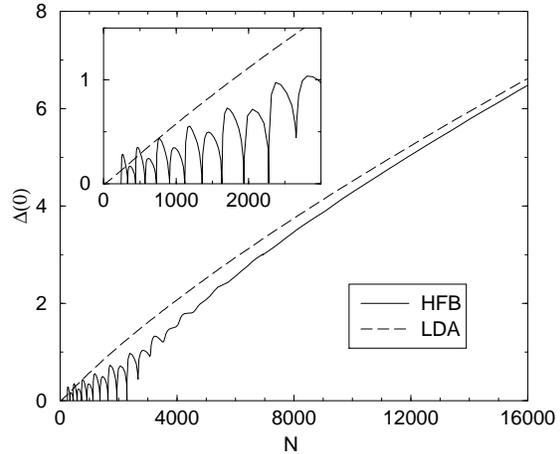,height=6.2cm}
\end{center}
\caption{Value of the pairing field in the center of the trap,
$\Delta(0)$ (in units of $\hbar \omega$), 
as a function of the number of particles, $N$, obtained
from HFB (solid line) and LDA (dashed line) calculations
[regularization method (b), cutoff $N_C = 100$, coupling constant $g =
-1$ in trap units].\label{fig4}}
\end{figure}

In \Fig{fig4} we present the HFB (full line) and LDA (dashed line)
results for the pairing field in the center of the trap, $\Delta(0)$,
as a function of the number of atoms $N$. The calculations are done
again at zero temperature and with a coupling constant $g = -1$ in
trap units. We observe that the two calculations are in reasonable
agreement for numbers of atoms greater than about $5000$, which
confirms the expectation that the LDA is a valid approximation for
systems with a large number of atoms.

What is particularly interesting to look at in this figure is the
region $N\lesssim 3000$. In this region the HFB results clearly show
the shell structure: the pairing field becomes zero for $N = 240,
330,440,\dots$, which are the harmonic oscillator ``magic
numbers''. One also realizes that the central value of the pairing
field is smaller if the outer shell corresponds to odd-parity states,
than in the case where the outer shell corresponds to even-parity
states. This can be understood easily, since the main contribution to
the pairing field comes from the states near the Fermi surface, and
only $s$ states can contribute to the pairing field at $r =
0$. Usually one expects that the LDA should at least reproduce the
value of the pairing field if the fluctuations due to shell effects
are averaged out, but our results show that the pairing field
calculated within the LDA is systematically too high. This might be
related to the fact that we are looking at the pairing field at one
particular point ($\vec{r} = 0$) rather than at the average gap at the
Fermi surface, as proposed in \Ref{Farine}.

When the number of atoms grows, above a value of about $2500$ the
shell structure starts to be washed out and gradually disappears due
to the stronger and stronger pairing correlations. This happens in the
region where the pairing field grows up to a value of about $\hbar
\omega$: when the pairing field becomes comparable with the oscillator
level spacing the pairing correlations in a closed shell system can
diffuse pairs of atoms towards the higher energy empty shell,
resulting in a non-zero pairing field. Globally, we observe that for
$N>5000$ the agreement between HFB and LDA is acceptable, even if the
LDA systematically overestimates the value of the pairing field at the
center.

Of course, the number of particles needed for the validity of the LDA
depends on the strength of the interaction; the true criterion which
has to be fulfilled reads $\Delta_{\mathit{LDA}} > \hbar\omega$. This
criterion can even be applied locally, as one can see in \Fig{fig2}:
there the HFB and LDA results are in perfect agreement except in the
region of $r\gtrsim 5.5 l_{ho}$, where $\Delta$ becomes smaller than
$\hbar \omega$.

\subsection{Results for non-zero temperature}
Now we will discuss some results for temperatures different from
zero. We are particularly interested in the following question: Within
the LDA, the critical temperature $T_C$ is different at each point
$\vec{r}$, i.e., when the temperature increases, the order parameter
vanishes at last in the center of the trap, where the local critical
temperature is the highest. In contrast to this, within the HFB
theory, the gap and the critical temperature are global properties,
and naively one would expect that, as long as the temperature is below
$T_C$, the pairing field extends over the whole volume of the
system. We will see that even in cases where the LDA works well at
zero temperature, it fails at non-zero temperature. On the other hand,
also the notion that the gap vanishes globally at $T = T_C$, has to be
revised in these cases.

\begin{figure}
\begin{center}
\epsfig{file=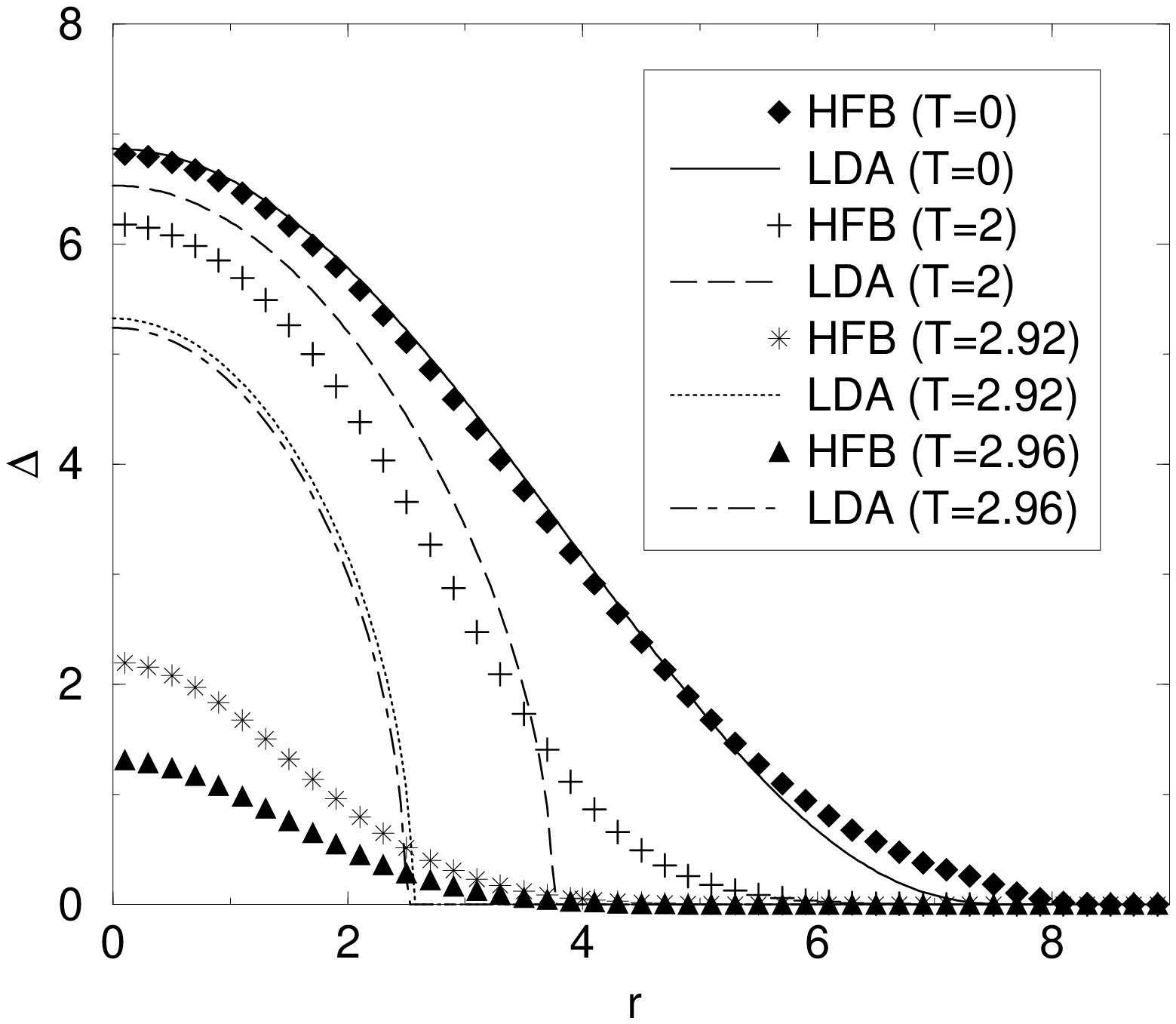,height=6.2cm}
\end{center}
\caption{Pairing field $\Delta$ (in units of $\hbar \omega$) as a
function of the distance $r$ (in units of $l_{ho}$) from the center of
the trap, for a chemical potential $\mu = 32\hbar \omega$,
corresponding to about $1.7\times 10^4$ atoms in the trap
[regularization method (b), cutoff $N_C = 100$, coupling constant $g =
-1$ in trap units]. Results obtained within numerical HFB calculations
(symbols) are compared with LDA results (lines) for different
temperatures $T$.\label{fig5}}
\begin{center}
\epsfig{file=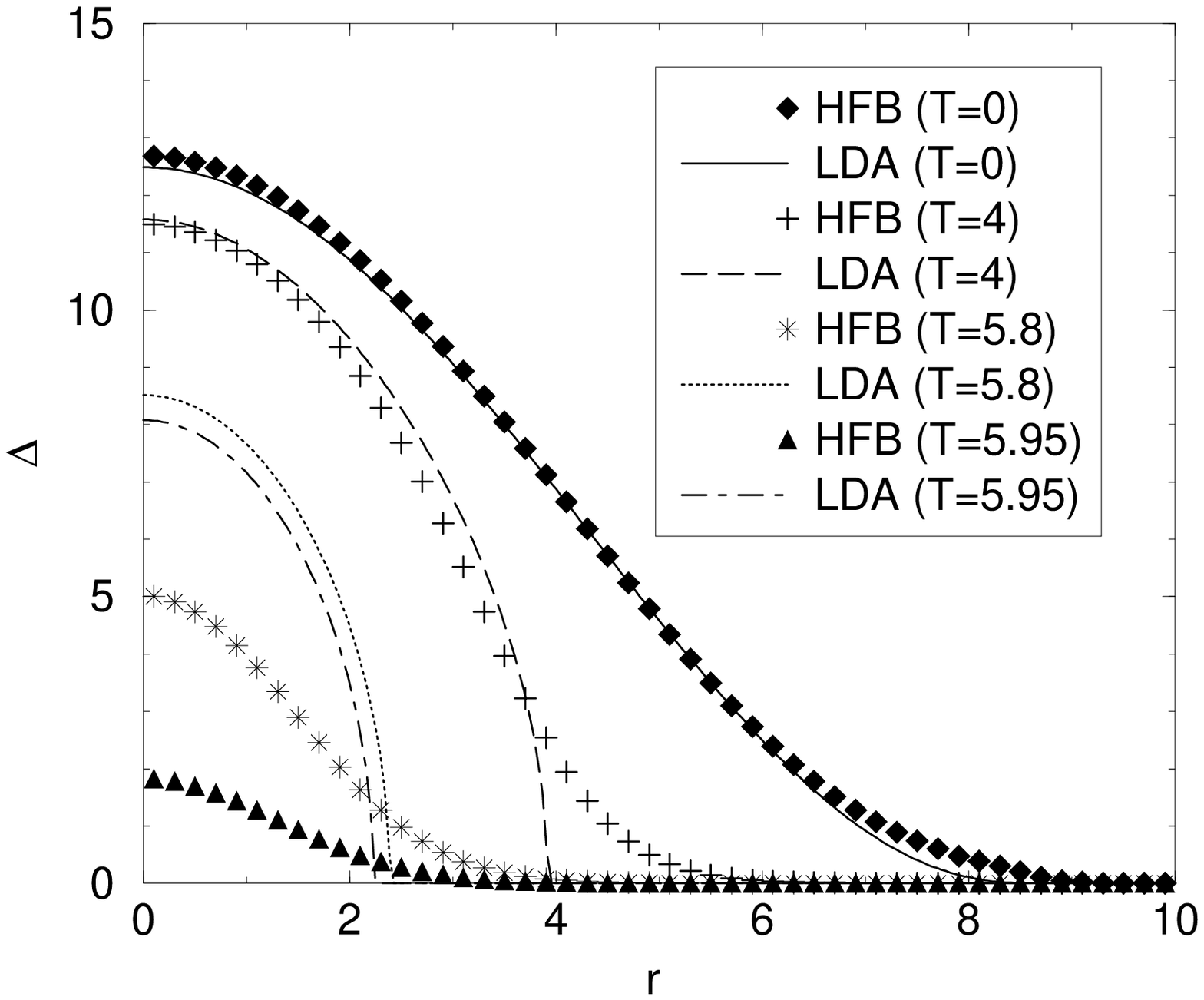,height=6.2cm}
\end{center}
\caption{Same as \Fig{fig5}, but for a chemical potential of $\mu =
40\hbar\omega$, corresponding to $N\approx 4\times10^4$ atoms in the
trap.\label{fig6}}
\end{figure}

In \Figs{fig5} and \ref{fig6} we show the HFB and LDA pairing fields
obtained at different temperatures, again for $g = -1$ (in trap units)
and regularization method (b). The chemical potentials chosen are $\mu
= 32 \hbar \omega$ in \Fig{fig5} and $\mu = 40 \hbar \omega$ in
\Fig{fig6}, corresponding to approximately $1.7 \times 10^4$ and
$4\times 10^4$ particles, respectively. We observe that the good
agreement obtained at zero temperature is deteriorated at higher
temperatures. In \Fig{fig5}, already at $T=2\hbar\omega/k_B$ the LDA
reproduces badly not only the tail of the pairing field profile, but
also the pairing field in the central region of the trap, in spite of
the fact that the pairing field is still large compared with
$\hbar\omega$ at this temperature. The LDA description gets worse and
worse for higher temperatures and results in an overestimation of the
central pairing field and in a too drastic cut of the queue of the
profile at large distances. Finally, the LDA method predicts a higher
critical temperature than the HFB one. We observed that $T_C$ is equal
to $3.89$ (in units of $\hbar\omega/k_B$) for LDA and to $2.98$ for
HFB. In \Fig{fig6}, the agreement is somewhat better. Since the
critical temperature is higher than in the previous case, the
agreement between LDA and HFB is maintained in a wider range of
temperatures. Up to $T = 4$ one can see that at least the central
region of the trap is well described by LDA. For higher temperatures,
we observe the same kind of deterioration of the LDA results shown in
\Fig{fig5}. Again, the critical temperature is higher in LDA ($7.08$)
than in HFB ($5.97$).

It is evident that the LDA does not correctly describe the phase
transition in both cases. On the other hand, also within the HFB
calculations one finds that with increasing temperature the pairing
field becomes more and more concentrated in the center of the
trap. Such a behavior has been predicted in \Ref{BaranovPetrov} using
the GL theory, the only assumption being that the
critical temperature is large compared with the trapping frequency,
$k_B T_C \gg \hbar\omega$. Let us briefly review the main results from
this theory and compare them with the results obtained from our
HFB calculations (the corresponding numbers are listed in
\Table{tab1}).
\begin{table}
\begin{center}
\begin{tabular}{c|c|c|c|c|c|c|c}
$\mu$ & $k_F(0)|a|$ & $T_C^{(0)}$ & $T_C$ & $\delta T_C$ &
$\delta T_C^{(GL)}$ & $l_\Delta$ & $l_{\Delta}^{(GL)}$\\
\hline
32 & 0.78 & 3.89 & 2.98 & 0.91 & 1.12 & 1.44 & 1.23\\
40 & 0.91 & 7.08 & 5.97 & 1.11 & 1.29 & 1.28 & 0.95
\end{tabular}
\end{center}
\caption{Comparison of results (in trap units) obtained from 
HFB calculations for the two cases $\mu = 32$ and $\mu = 40$ shown in
\Figs{fig5} and \ref{fig6} [coupling constant $g = 1$ in trap units,
regularization method (b), $N_C = 100$] and the corresponding results
obtained from the GL theory.}
\label{tab1}
\end{table}

In the GL theory the critical temperature $T_C$ is predicted to be lower than the
critical temperature $T_C^{(0)}$ obtained from the LDA. The difference
can be written as
\begin{align}
\delta T_C =& T_C^{(0)}-T_C\nonumber\\
=& \frac{3\hbar\Omega}{k_B}
  \sqrt{\frac{7\zeta(3)}{48\pi^2}\Big(1+\frac{\pi}{4k_F(0)|a|}\Big)}\,,
\label{tcgl}
\end{align}
where $\zeta$ denotes the Riemann zeta function ($\zeta(3) =
1.202\dots$). In the derivation of \Eq{tcgl} in \Ref{BaranovPetrov}
the Hartree potential has been neglected. Here we will include the
Hartree potential by using an effective oscillator frequency $\Omega >
\omega$. Since near $T_C$ the pairing field is concentrated in the
center of the trap, we define $\Omega$ by expanding the potential
around $r = 0$:
\begin{equation}
\Omega = m \sqrt{\vec{\nabla}^2
  [U_0(\vec{r})+W(\vec{r})\big)]_{\vec{r} = 0}}\,.
\end{equation}
Within the Thomas-Fermi approximation for the density profile the
effective oscillator frequency can be written as
\begin{equation}
\Omega = \frac{\omega}{1-\frac{2k_F(0)|a|}{\pi}}\,.
\label{omeff}
\end{equation}
The estimates for $\delta T_C$ obtained by inserting the numerical
values for $k_F(0)|a|$ given in \Table{tab1} into \Eqs{tcgl} and
(\ref{omeff}) are very reasonable. This can be seen by comparing them
with the $\delta T_C$ values obtained from the HFB calculations, which
are also listed in \Table{tab1}. If one considers that these numbers
can only be a rough estimate, since $k_B T_C$ is not really very large
compared with $\hbar\Omega$, the agreement with the HFB results is
very satisfying.

Not only the critical temperature, also the shape of the order
parameter near the critical temperature can be obtained from the GL
theory. It can be shown that for temperatures very close to $T_C$ the
pairing field has the form of a Gaussian,
\begin{equation}
\Delta(\vec{r}) = \Delta(0) \exp\Big(-\frac{\vec{r}^2}{2l_\Delta^2}\Big)\,.
\end{equation}
In contrast to the LDA result, the radius $l_\Delta$ of this Gaussian
is predicted to stay finite in the limit $T\to T_C$, as it is the case for
the solution of the HFB equations. Its value is
given by
\begin{equation}
l_\Delta^2 = R_{TF}^2 \frac{\hbar\Omega}{k_B T}
  \sqrt{\frac{7\zeta(3)}{48\pi^2}\,\frac{1}{1+\frac{\pi}{4k_F(0)|a|}}}
\,.
\label{rgl}
\end{equation}
In \Ref{BaranovPetrov} the quantity $R_{TF}$ was defined as the
Thomas-Fermi radius of the cloud, $R_{TF} =
\sqrt{2\mu/(m\omega^2)}$. Generalizing the derivation of \Eq{rgl} to
the case of a non-vanishing Hartree field, we see that the
corresponding parameter for the pairing field near the center of the
trap is given by
\begin{align}
R_{TF} \to\, &\sqrt{\frac{2[\mu-W(0)]}{m\Omega^2}}\nonumber\\
&= \Big(1-\frac{2k_F(0)|a|}{\pi}\Big)k_F(0) l_{ho}^2\,.
\label{rtf}
\end{align}
On the other hand, the HFB pairing fields corresponding to the
temperatures next to $T_C$ shown in \Figs{fig5} and \ref{fig6} are
also perfectly fitted by Gaussians. As shown in \Table{tab1}, the
agreement between the radii obtained from this fit are again in
reasonable agreement with the radii obtained from \Eqs{rgl} and
(\ref{rtf}). The deviations are of the order of $30\%$, which is even
better than one could have expected, since the parameter
$\hbar\Omega/(k_B T_C)$ is not very small in the present case.

\begin{figure}
\begin{center}
\epsfig{file=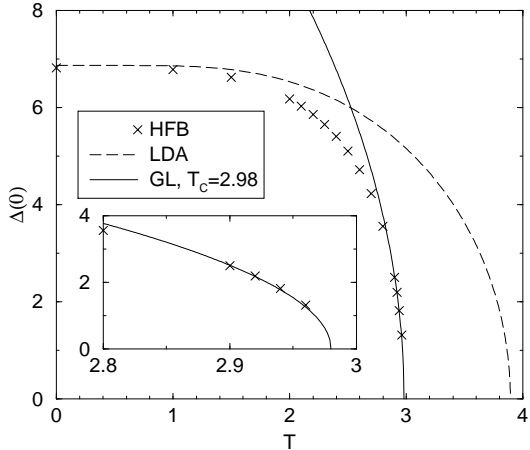,height=6.2cm}
\end{center}
\caption{Value of the pairing field in the center of the trap,
$\Delta(0)$ (in units of $\hbar \omega$), as a function of temperature
$T$ (in units of $\hbar\omega/k_B$) for a chemical potential $\mu =
32\hbar \omega$, corresponding to about $1.7\times 10^4$ atoms in the
trap [regularization method (b), cutoff $N_C = 100$, coupling constant
$g = -1$ in trap units]. Results obtained within numerical HFB
calculations (symbols) are compared with the LDA result (dashed line)
and with the formula (\ref{deltagl}) obtained from the GL theory
(solid line).\label{fig7}}
\end{figure}

\begin{figure}
\begin{center}
\epsfig{file=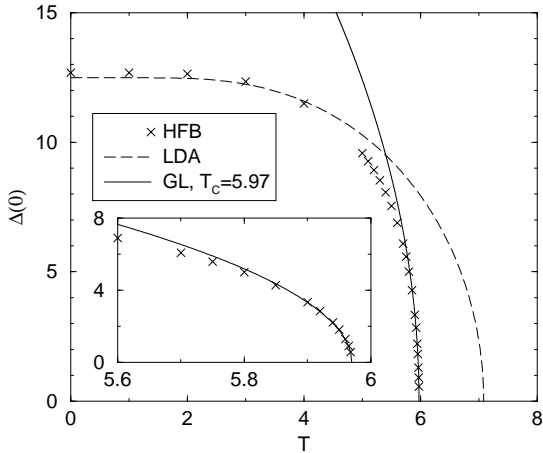,height=6.2cm}
\end{center}
\caption{Same as \Fig{fig7}, but for a chemical potential of $\mu =
40\hbar\omega$, corresponding to $N\approx 4\times10^4$ atoms in the
trap.\label{fig8}}
\end{figure}

Finally, let us look more closely at the critical behavior near
$T_C$. Again, from the GL theory one can derive that for $T\to T_C$
the value of the pairing field in the center should go to zero like
\begin{equation}
\Delta(0) = \sqrt{\frac{16\pi^2\sqrt{2}}{7\zeta(3)}\,T_C(T_C-T)}\,.
\label{deltagl}
\end{equation}
As shown in \Figs{fig7} and \ref{fig8}, this formula is very well
satisfied by the HFB results in both cases, $\mu = 32$ and
$\mu = 40$ (in trap units). Note that the prefactor in \Eq{deltagl}
differs from the prefactor in LDA. In LDA one finds for $T\approx
T_C^{(0)}$
\begin{equation}
\Delta_{\mathit{LDA}}(0) =
  \sqrt{\frac{8\pi^2}{7\zeta(3)}\,T_C^{(0)}(T_C^{(0)}-T)}\,.
\end{equation}
The different prefactor, as well as the different critical temperature
and the finite radius of the pairing field, are due to the ``kinetic''
term $\propto \Delta\vec{\nabla}^2 \Delta$ in the GL energy
functional, which is absent in the LDA and which is very important for
the description of the strongly $\vec{r}$ dependent pairing field near
the critical temperature.

As a final remark let us mention that the different calculations which
we have compared in this paper, are all based on mean-field theory,
and therefore do not take into account fluctuations of the order
parameter $\Delta$. It is well-known that fluctuations are very
important near the phase transition, and in particular in a situation
where $k_F |a|$ is not small, as it is the case here, they can lead to
a considerable change of the critical temperature. Anyway, what we
wanted to point out here, is that the LDA gives the wrong $T_C$ as
compared with a theory taking into account the inhomogeneity of the
system. From this result we conclude that in order to have a reliable
prediction of $T_C$ for the trapped system, it is not sufficient to do
a reliable calculation of $T_C$ (even including fluctuations) for a
homogeneous gas and then apply the LDA.

\section{Conclusions}
\label{secconclusions}
In this paper we have shown a detailed comparison between HFB and LDA
calculations at $T=0$ and at $T\neq 0$ for a low density gas of
superfluid fermionic atoms trapped by a spherical harmonic potential.
We have used a zero-range interaction for the atoms and we have
proposed an improvement of the regularization method adopted to remove
the ultraviolet divergence \cite{brca}. This improvement is a
modification of a procedure proposed for nuclear systems in
\Ref{bul}, where the Thomas-Fermi approximation is used in the
calculation of the regular part of the Green's function
$G^{0\,\mathit{reg}}_\mu$, \Eq{gregtf}. The use of the Thomas-Fermi
approximation allows to treat systems with a large number of atoms
much easier than in the calculations of \Ref{brca}. On the other
hand, our modification considerably improves the convergence rate of
the procedure with respect to the numerical cutoff. By using this
regularization method we have observed that the LDA results are in
quite good agreement with the corresponding HFB results at zero
temperature and for systems with a relatively large number of atoms,
where the shell structure effects are washed out. The shell effects,
which are important for small systems where the pairing field is
smaller than the harmonic level spacing $\hbar \omega$, cannot
obviously be reproduced by a LDA calculation.

For non-zero temperatures the agreement between HFB and LDA is
deteriorated even in those cases where it was good at $T=0$.  In
general, LDA overestimates the value of the pairing field in the
center of the trap, cuts too drastically the tail of the radial
profile of the pairing field at large distances, and overestimates the
critical temperature with respect to HFB. We have verified that this
discrepancy between the HFB and LDA results at $T$ different from zero
can be nicely predicted by using the GL theory \cite{BaranovPetrov} in
cases where the critical temperature is much larger than the harmonic
level spacing.

In this article we considered only spherical traps. However, the traps
used in experiments are usually cigar-shaped with a low longitudinal
and a high transverse trapping frequency, $\omega_z \ll
\omega_\perp$. In this case it is possible that the pairing field,
even if it is larger than $\hbar\omega_z$, is still smaller than
$\hbar\omega_\perp$, and the LDA would probably not work. Therefore in
principle one should also perform deformed HFB calculations, but at
the moment this seems to be numerically very difficult. On the other
hand, as noted above, even in the case where $\Delta$ is large
compared with both trapping frequencies, the LDA is not adequate at
non-zero temperature. Therefore a first step to study non-spherical
traps could be to generalize the GL theory of \Ref{BaranovPetrov} to
the deformed case.
\\[5mm]
\begin{acknowledgments}
The authors wish to thank Elias Khan, Peter Schuck and Nguyen Van Giai
for useful discussions. We thank Yvan Castin for supplying us the code
for the numerical calculation of the Green's function
$G_\mu^{0\,\mathit{reg}}$ which we used to check the TF
approximation. M.G. is a recipient of a European Community Marie Curie
Fellowship, M.U. acknowledges support by the Alexander von Humboldt
foundation.
\end{acknowledgments}

\end{document}